\documentclass[12pt]{article}
\usepackage{graphicx}

\usepackage{amsfonts}
\usepackage{amssymb}
\usepackage{amsthm}
\usepackage{amsmath}
\begin{document}
\begin{titlepage}
\begin{flushright}
hep-ph/0311183{\hskip.5cm}
\end{flushright}
\begin{centering}
\vspace{.3in}
{\bf GUT Precursors in $SU(3)^3$-Type Model and $N_{COLOUR}>3$}\\
\vspace{2 cm}
{N.D. Tracas} \\
\vspace{.3cm}
{\it {Physics
Department, National Technical University,\\
       Athens 157 73 , Greece}}\\
\end{centering}
\vspace{.1in}

\abstract{\noindent
We investigate  the $SU(3)^3$ GUT model when
signs of the model (precursors),  due to low compactification scale,
appear before the gauge couplings of the
Standard Model get unified.
The Kaluza-Klein state contribution seems to lead
the gauge couplings to unification through a wide energy scale only
in the case when the colour group is augmented to $SU(4)$.}
\vfill
\hrule width 6.7cm
\begin{flushleft}
November 2003
\end{flushleft}
\end{titlepage}
\vspace{.5cm}
\noindent
{\bf Introduction}\\
The unification of the SM gauge couplings to a common value at
some high energy scale still plays an attractive role in our
efforts to understand the fundamental interactions of nature. The
supersymmetric extensions of the SM provides us with such a
scenario. The energy scale of this unification is,however, of the
order of $10^{16}$GeV, rendering experimental evidence at best
indirect. On the other hand, large extra dimension can lower the
scale of gauge coupling unification due to the appearance of the
Kaluza-Klein (KK) tower of states above the compactification scale%
\cite{low_unif,low_unif1,low_unif2,low_unif3,low_unif4,low_unif5}%
.
Inclusion of KK states (either from the higgs and the gauge bosons or from
all the spectrum) in the MSSM could lead to lower energy scale unification%
\cite{low_unif1,tracas_leontaris}%
.
The same idea was also applied%
\cite{tracas_leontaris} to successful GUT models that could be
originated from strings, namely the $S(4)\times SU(2)_L\times
SU(2)_R$ and the $SU(3)^3$ models. In both models, the inclusion
of several numbers of exotic states (i.e. states not appearing in
the SM spectrum but present in models derived from strings) helps
in providing the gauge coupling unification.
Applying the standard Higgs mechanism to break the GUT model
obviously the masses of the extra GUT fields are of the order of
the vev used which in turn is of the order of the GUT breaking
scale. On the other hand, using the orbifold method to break the
larger symmetry, the extra fields acquire masses of the order of
the compactification scale (the inverse of the circle $S^1$ for
one extra dimension). Thus, in case that this scale is lower than
the scale where the gauge coupling meet ($M_{GUT}$), we have the
appearance of KK states of the extra GUT fields. These fields are
the so called ``precursors" which of course influence the running
of the $\beta$-functions for energy scales above the
compactification one.
In the present work, we will concentrate in the $SU(3)^3$ GUT model.
Assuming a low compactification scale we will try to accomplish
the gauge coupling unification incorporating in the MSSM
$\beta$-functions the contribution of the GUT precursors.

\vspace{.5cm}
\noindent
{\bf The Model}\\
Let us briefly describe the $SU(3)_C\times SU(3)_L\times SU(3)_R$
GUT model, which is one of the few that can be derived from
strings. The MSSM content can be found in the $\underline{27}$
representation of the $E_6$:
\[
27\rightarrow (3,\overline{3},1)+(\overline{3},1,3)+(1,3,\overline{3})
\]
where
\begin{equation}
\label{SU3}
\begin{split}
&Q=(3,\overline{3},1)=
\left(
\begin{array}{c}
u\\
d\\
D
\end{array}
\right),
\quad
Q^c=(\overline{3},1,3)=
\left(
\begin{array}{c}
u^c\\
d^c\\
D^c
\end{array}
\right)
\\
&L=(1,3,\overline{3})=
\left(
\begin{array}{ccc}
h_0  &h^+  &e^c\\
h^-  &\overline{h}_0  & \nu^c\\
e    & \nu  &N
\end{array}
\right)
\end{split}
\end{equation}
The emergence of the SM comes as follows: The $SU(3)_C$ is the
colour group. The second $SU(3)$ breaks to $SU(2)_L\times U(1)_L$
while $SU(3)_R$ breaks to $U(1)_R$. The SM $U(1)$ comes as a
linear combination of the two $U(1)_{L,R}$. The hypercharge $Y$ is
related to the $X$ and $Æ$ charges of the $U(1)_L$ and $U(1)_R$
correspondingly, through the relation
\begin{equation}
\label{Y}
Y=\frac{1}{\sqrt{5}}\,X+\frac{2}{\sqrt{5}}\,Z
\end{equation}
while the corresponding relation between the couplings at the breaking scale
is
\[
\alpha_3=\alpha_C,\quad
\alpha_2=\alpha_L,\quad
\alpha_Y^{-1}=\frac15 \alpha_L^{-1}+\frac45 \alpha_R^{-1}
\]

The one-loop $\beta$-functions are given by
\begin{equation}
\label{beta_fun}
\begin{split}
&\beta_C=-9+\frac12\left(3n_Q+3n_{Q^c}\right)\\
&\beta_L=-9+\frac12\left(3n_Q+3n_{L}\right)\\
&\beta_R=-9+\frac12\left(3n_{Q^c}+3n_{L}\right)
\end{split}
\end{equation}
where $n$ shows the number of the corresponding representation.

\vspace{.5cm}
\noindent
{\bf The Precursor Contribution to the MSSM $\beta$-Functions}
\\
The general form for the Kaluza-Klein state contribution to the
$\beta$-function is:
\begin{equation}
\label{KK_b_function}
\beta_{KK}=-2C_2(G)+\sum_i T(R_i)
\end{equation}
where the first term comes from the gauge multiplet (gauge bosons
and gauginos), while the second comes from the chiral multiplets
(quarks, leptons, higgs and superpartners).
We should find the contribution to the MSSM $\beta$-functions coming
from the members of the gauge sector and chiral sector which do not appear
in the corresponding MSSM sectors.
Let us start from the former ones. In the $SU(3)_L\rightarrow SU(2)_L\times U(1)$
breaking, the adjoint of the $SU(3)$ gives:
\[
8\rightarrow 3\,+\,2\,+\,\bar{2}\,+\,1
\]
while the $U(1)$ charges are zero for the 3 and the singlet and
$\pm\sqrt{3}/2$ for the 2's. The 3 are the gauge bosons of the
$SU(2)_L$ of the SM, while the singlet is completely blind in all
SM interactions. The doublets appear as spin 1 (plus the SUSY
partners) particles having $SU(2)_L$ gauge interactions.

Now, the $C_2(G)$, appearing in Eq.(\ref{KK_b_function}), for an $SU(N)$ group
is equal to N which comes from the contraction of the structure constants of
the group: $f_{ijk}f_{i'jk}=N\delta_{ii'}$. In our case we should find this
summation when $i$, $i'$ and $j$ correspond to the doublet while $k$ corresponds
to the triplet. In the $SU(3)$ case this summation gives 3/2 instead of 3\footnote{
For the general case $SU(N+1)\rightarrow SU(N)\,+\, U(1)$, this summation gives
$(N^2-1)/N$.}.
Therefore the contribution of these two doublets to the $\beta$-function of
the $SU(2)$ is $-2*(3/2)=-3$ for each doublet.

Let us now find the contribution of the doublets to the SM U(1) $\beta$-function.
The charge under $U(1)_L$ is $\sqrt{3}/2$. Their charges under the $U(1)_R$,
coming from the breaking of the $SU(3)_R$, is of course zero. Therefore, using
Eq.(\ref{Y}), the contribution of each doublet to the $U(1)$ $\beta$-function
is: $-2[(1/\sqrt{5})(\sqrt{3}/2)]^2*2$.

Following the same procedure, the contribution of each doublet, coming
from the breaking of the adjoint of the $SU(3)_R$, to the SM $U(1)$ is:
$-2[(2/\sqrt{5})(\sqrt{3}/2)]^2*2$.

We turn now to the contribution of the $D$ and $D^c$ appearing in the
$Q$ and $Q^c$ representations. Both, $D$ and $D^c$, being in the
fundamental representation of the $SU(3)_C$, contribute a term $(1/2)N_g$
each, in the colour group $\beta$-function ($N_g$ is the number of generations).
In the breaking of $SU(3)_L$, the $D$'s appear as the singlets in the
breaking of the fundamental representation: $3\rightarrow 2\,+\,1$, with
charge under $U(1)_L$ equal to $-1/\sqrt{3}$. Therefore, they do not
contribute to the $SU(2)_L$ $\beta$-function while their contribution to
the $U(1)$ $\beta$-function is $[(1/\sqrt{5})(-1/\sqrt{3})]^2 3N_g=(1/5)N_g$.
Finally, the $D^c$ do not contribute to the $SU(2)_L$ $\beta$-function and
since $SU(3)_R$ breaks to $U(1)_R$, the charge under the last group could
be arbitrary, fixed in such a way as to give the correct electrical charge.
Indeed, the $U(1)_R$ charge of the $D^c$ is $1/(2\sqrt{3})$ and the
contribution to the $U(1)$ $\beta$-function is
$[(2/\sqrt{5})(1/(2\sqrt{3}))]^2*3N_g=(1/5)N_g$.
The two states, $N$ and $\nu^c$, appearing in $L$ are totally blind in the
SM interactions.

Gathering all the above we can write the contribution of the Kaluza-Klein
states of the non-SM particles to the $\beta$-function as follows ($N_g=3$):
\begin{equation}
\label{extra_part}
\beta_3=3,\quad\quad
\beta_2=-6,\quad\quad
\beta_Y=-24/5
\end{equation}

\vspace{.5cm}
\noindent
{\bf The Running of the One-Loop $\beta$-Functions}
\\
We assume that from $M_Z$ to $M_{SUSY}=1$TeV, we have the non-SUSY SM
$\beta$-functions. We use the following experimental values as our
starting point at $M_Z$:
\[
\sin^2\theta_W=0.23151\pm .00017,\quad\quad
\alpha_{em}=1/128.9,\quad\quad
\alpha_s=0.119\pm 0.003
\]
From $M_{SUSY}$ to the compactification scale $M_C$ we have
the MSSM $\beta$-functions. Above $M_C$, all Kaluza-Klein states
start to appear. We assume the successful approximation of
incorporating the massive KK-states with masses less than the
running scale. The running of the couplings above $M_C$ is given by
\[
\alpha_i^{-1}(M')=\alpha_i^{-1}(M_C)-
\frac{\beta_i}{2\pi}\left(2N\log\frac{M'}{M_C}-2\log(N!)\right)
\]
where N is an integer such that $(N+1)M_C>M'>NM_C$, which counts
the KK-states that have masses below the running scale (we have assumed
only one extra dimension  and in that case the multiplicity of the states
at each level is 2).
\begin{figure}[!t]
\centering
\includegraphics[scale=0.6]{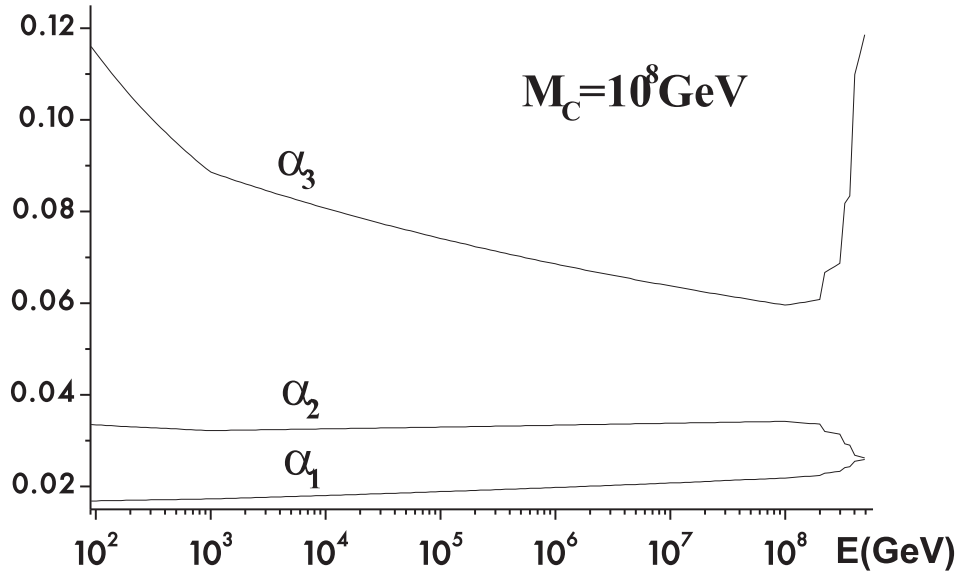}
\caption{Running of the SM couplings. $M_C=10^8$GeV.}
\label{alphasN3}
\end{figure}
In Fig.(\ref{alphasN3}) we show such a running. It is obvious from
the positivity of the KK-state contribution to the
$\beta$-function that the three couplings could not converge to a
point. And this fact, of course, persists whatever is the choice
of the scale $M_C$, since we know that the MSSM couplings converge
at the scale $~10^{16}$GeV which means that the strong coupling is
always larger than the other two up to that scale.

\vspace{1cm}
\noindent
{\bf Upgrading the Colour Group to $SU(N)$, $N>3$}
\\
The idea that at high energies the colour group is $SU(N>3)$ has
a long history%
\cite{Hooft}%
and was considered as requirement for ``asymptotic convergence" on top
of asymptotic freedom. Recently, this idea was applied to Grand Unification%
\cite{Branco}%
.
In our case, therefore, it is more that tempting to investigate the case
where the colour group of our model is upgraded to $SU(N)$ with $N>3$.
We assume that the breaking from $SU(N)_C$ down to $SU(3)_C$ is stepwise:
$SU(N)_C\rightarrow SU(N-1)_C\,+\,U(1)$. The
conjugate and the fundamental representation breaking are:
\begin{equation}
\label{SUN}
\begin{split}
(N^2-1)\, & \,\longrightarrow ((N-1)^2-1)\, +\, (N-1)\, +\, \overline{(N-1)}\, +\, 1\\
 N \, & \,\longrightarrow (N-1)\, +\, 1
\end{split}
\end{equation}
We further assume that all singlets produced by these breakings
get masses and therefore they do not contribute to the colour group.
At the end of the day, we are left with the fundamental representation
of $SU(3)_C$ coming from the fundamental one of $SU(N)$ (i.e. the
coloured quarks of the SM) while from the conjugate representation of
$SU(N)$ we get the gluons plus a number of $3$'s and $\overline{3}$'s.
The number of these states is proportional to $N-3$ (the number of
times breaking occurs from $SU(N)_C$ to $SU(3)_C$). For example,
if we have an $SU(5)_C$ group, the breaking of the adjoint down to
$SU(3)_C$  is:
\[
\begin{array}{ccc}
24   & \longrightarrow &15\,\,+\,\,4\,+\,\overline{4}\,+\,1   \\
     &                 &
     \begin{array}{lc}
     \,\,\,\,\,\,\,\Big\downarrow  &
         \begin{array}{lc}
         \,\,\,\Big\downarrow  &
             \begin{array}{lc}
             \Big\downarrow  &     \\
             \multicolumn{2}{l}{\overline{3}\,+\,1}
             \end{array}\\
         \multicolumn{2}{l}{\,\,\,3\,+\,1}
         \end{array}\\
     \multicolumn{2}{l}{\,\,\,\,\,\,\,8\,\,+\,\,3\,+\,\overline{3}\,+1}
     \end{array}
\end{array}
\]
The contribution of the $3$'s and $\overline{3}$'s (KK-) states to the
$\beta$-function of $SU(3)_C$ will be:
\[
-2 \, \frac{N(N-2)}{(N-1)}\, 2\, (N-3)
\]

\begin{figure}[!t]
\centering
\includegraphics[scale=0.6]{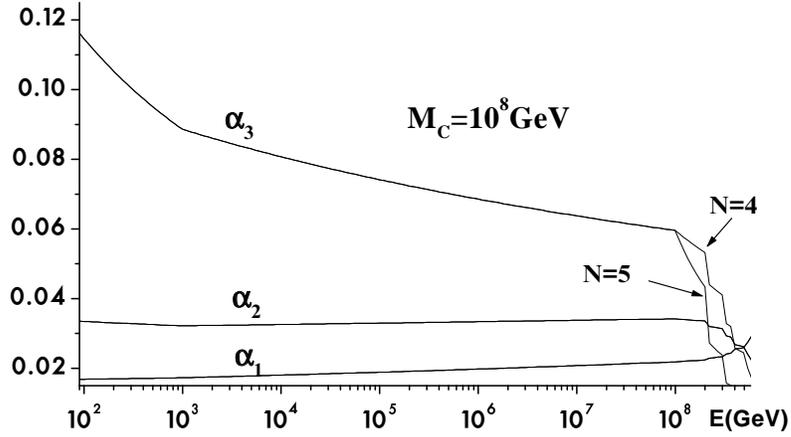}
\caption{Running of the SM couplings for $N=4$ and $N=5$.
$M_C=10^8$GeV.}
\label{alphasN4N5}
\end{figure}

Now we are ready to run the coupling constants with the above new contribution.
In Fig(\ref{alphasN4N5}) we show the running for two values of
$N=4$ and $5$. We clearly see that for $N=4$ the three coupling
constants unify to a good accuracy while for $N=5$ the strong
coupling decreases very rapidly and unification is missed. Of
course, for higher values of $N$ the situation will be even worst
(thus strong $\beta$-function will be even more negative).

\begin{figure}[!b]
\centering
\includegraphics[scale=0.55]{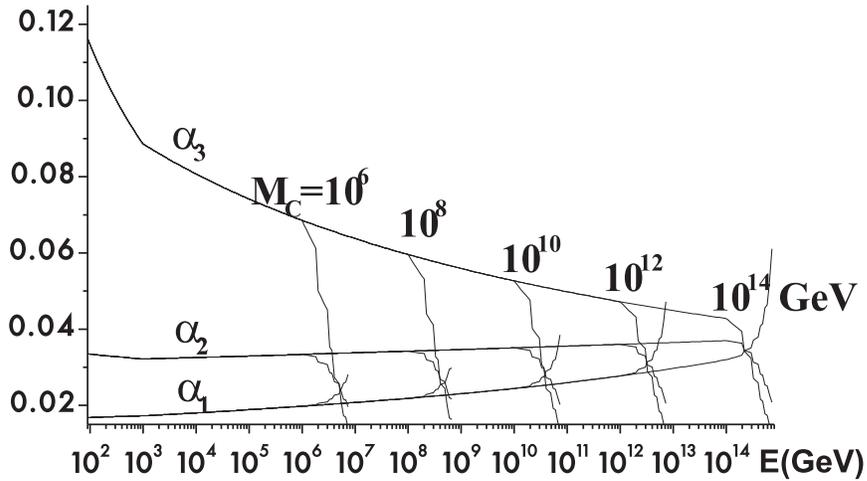}
\caption{Running of the SM couplings for $N=4$ and several values
of the compactification scale $M_C=10^6,10^{8},10^{10},10^{12}$ and
$10^{14}$GeV.}
\label{alphasN4}
\end{figure}

In Fig(\ref{alphasN4}) we show the running for $N=4$ and for
several values of the compactification scale $M_C$. We see that
the unification is independent of the chosen compactification
scale. Of course, the unification scale is just above the
compactification one and only a small number of the KK-states
contribute to the running. Nevertheless, we can achieve a low
energy unification and the value of the unified coupling is well
in the perturbative region.

\vspace{1cm}
\noindent
{\bf Conclusions}
\\
Using the precursor idea, i.e. the appearance of KK states of the
$SU(3)^3$ GUT model before the SM couplings get unified, we have
study the gauge coupling running. The contribution of KK states,
due to low compactification scale, accelerates the convergence of
the couplings but unification seems to accomplished only if the
colour group is $SU(4)$ at the GUT scale providing therefore extra
precursors. The unification scale appears near the chosen
compactification scale but the unification itself is independent
of the latter scale.\\


We would like to thank A. Kehagias for helpful discussions.


\begin{thebibliography}{99}
\bibitem{low_unif}
I. Antoniadis, Phys. Lett. B246 (1990) 377;
I. Antoniadis, K. Benakli and M. Quiros, Phys. Lett. B331 (1994)313.
\bibitem{low_unif1}
K. Dienes, E.Dudas and T.Gherghetta, Phys. Lett. B436 (1998) 55;
Nucl. Phys. B537 (1999) 47; arXiv:hep-ph/9807522; arXiv:hep-ph/0210294.
\bibitem{low_unif2}
D. Ghilencea and G.G. Ross, Phys. Lett. B442 (1998) 165.
\bibitem{low_unif3}
K. Benakli, Phys.Rev. D60 (1999) 104002
\bibitem{low_unif4}
C.P. Burges, L.E. Iban\~{e}z and F. Quevedo, Phys.Lett. B447 (1999) 257;
L.E. Iban\~{e}z, C. Mu\~{n}oz and S. Rigolin, Nucl.Phys. B553 (1999) 43;
T. Li, arXiv:hep-ph/9903371.
\bibitem{low_unif5}
A. Delgado and M. Quiros, Nucl.Phys. B559 (1999) 235.
\bibitem{tracas_leontaris}
G.K. Leontaris and N.D. Tracas, Phys. Lett. B470 (1999) 84.
\bibitem{Hooft}
G. 't Hooft, Phys.Lett. B{\bf 109}(1982)1848.
\bibitem{Branco}
G.C. Branco, J.-M. G\'erard, R. Gonz\'alez Felipe
and B.M. Nobre, arXiv:hepph/0305092.%

\end{thebibliography}
\end{document}